\documentclass[prl,letterpaper,twocolumn,aps,epsf]{revtex4}
\usepackage{dcolumn}
\usepackage{amsmath}
\usepackage{graphicx}

\newcommand{\ve}[1]{\mathbf{#1}}

\begin{document}
\title{Thermal Fluctuations of Vortex Matter in Trapped Bose--Einstein
  Condensates}

\author{S. Kragset${}^1$, E. Babaev${}^{2,1}$, and A. Sudb\o ${}^1$}

\affiliation{ ${}^1$ Department of Physics, Norwegian University of
  Science and Technology,
  N-7491 Trondheim, Norway\\
  ${}^2$ Laboratory of Atomic and Solid State Physics, Cornell
  University, Ithaca, NY 14853-2501, USA }

\begin{abstract}
  We perform Monte Carlo studies of vortices in three dimensions in a
  cylindrical confinement, with uniform and nonuniform density. The
  former is relevant to rotating ${}^4$He, the latter is relevant to a
  rotating trapped Bose--Einstein condensate. In the former case we
  find dominant angular thermal vortex fluctuations close to the
  cylinder wall. For the latter case, a novel effect is that at low
  temperatures the vortex solid close to the center of the trap
  crosses directly over to a tension-less vortex tangle near the edge
  of the trap.  At higher temperatures an intermediate tensionful
  vortex liquid located between the vortex solid and the vortex
  tangle, may exist.
\end{abstract}

\maketitle 

Topological vortex excitations are hallmarks of quantum ordered states
such as superconductors and superfluids. Vortices are important at
vastly different length scales ranging from the dynamics of neutron
stars to transport properties in superconductors and rotational
response of ultracold atomic gases in optical traps
\cite{exp,Ketterle}.  Particularly rich physics is associated with
various orderings of quantum vortices, and the transitions between
them. In certain systems, even numerous aggregate states of vortex
matter are possible \cite{Nature_2004}. The simplest, yet extremely
important, temperature-induced phase transition in vortex matter in a
one-component superconductor or superfluid is the one from an ordered
vortex line lattice (VLL) to a disordered vortex liquid (VL) state.
This has been a subject of intense research in the context of
high-$T_c$ superconductivity.

Remarkable progress has recently been made in the physics of vortices
in Bose-Einstein condensates (BEC) of ultra-cold atoms
\cite{exp,Ketterle}. These systems are extremely clean and have
physical parameters controllable in a wide range.  This has led to
many suggestions for testing a number of general physical concepts in
BECs. Since individual vortices, as well as their ordering patterns,
can be resolved in ultra-cold gases \cite{exp,Ketterle}, a natural
question arises if in these systems a thermally induced phase
transition from VLL to VL can be observed, and if it can shed more
light on fundamental properties of the VL states in quantum fluids in
general. Detailed theoretical studies of a model uniform system, with
density corresponding to the values attainable in the center of a
trap, indicate that in present experiments rotation is too slow and/or
particle numbers are too high to obtain a VLL melting if the
approximation of a uniform system holds \cite{baym}.  We also mention
studies of quantum VLL melting in a one dimensional optical lattice
\cite{1d}. In a harmonic trap, the condensate density will gradually
be depleted from the center of the trap towards its edge.  This
suggests that thermal fluctuation effects will be enhanced close to
the edges. We thus focus on investigating possible crossover states of
vortex matter in a spatially inhomogeneous system such as a trapped
BEC.

Melting of vortex lines in a bosonic condensate can be modeled by the
uniformly frustrated lattice 3D $XY$ model (see e.g.
\cite{hetzel,tachiki,sudbo}). We study a system which takes into
account the presence of a harmonic trap by using the Hamiltonian
\begin{equation}
  \label{traphamilton}
  H = - \sum_{\langle ij \rangle} P_{ij} \cos(\theta_j -
  \theta_i - A_{ij}),
\end{equation}
where $\theta_i$ is the phase of the condensate wave function at
position $i$. The factor $P_{ij} \equiv P(r_{ij}) =1-(r_{ij}/R)^2 \
{\rm if} \ r_{ij} \leq R$, where $R$ is the cloud size and $r_{ij}$ is
the radial distance from the cloud center. For $r_{ij} > R$ we set
$P_{ij}=0$. We account for circulation by introducing the potential
$A_{ij}= \int_i^j d {\bf l} \cdot {\bf A}$. Here, $\nabla \times
\ve{A} = (0, 0, 2 \pi f)$, and $f$ is the number of rotation-induced
vortices per plaquette in the $xy$-plane.

We perform Monte Carlo (MC) simulations on Eq. (\ref{traphamilton})
with the Metropolis algorithm. We have considered cubic numerical
grids of size $L^3$, with $L=36,72$, and along the $z$-direction we
impose periodic boundary conditions to model an elongated system. The
filling fraction is $f=1/36$, and temperatures $T \in (0.30, 3.0)$
\cite{details}. The temperature at which the Bose condensation takes
place at zero rotation in a bulk system, is given by $T=2.2$ in our
units. We note that a rapidly rotating system should be better
described by the Lowest Landau Level (LLL) approximation
\cite{watanabe-cooper}, characterized for instance by the fact that
the intervortex separation is comparable to the vortex-core size.
However, several experiments on elongated systems are clearly outside
the LLL regime. For instance, in Ref.  \cite{Ketterle}, the
inter-vortex separation is given as $5.0 \mu m$, while the healing
length (vortex-core radius $\xi$) is $0.2 \mu m$. A parameter
estimating the validity of the LLL approximation is the ratio
$\lambda$ of the interaction energy scale to the level spacing of the
transverse harmonic confinement \cite{watanabe-cooper}, $\lambda = 4
\pi \hbar^2 a_s n/(M \hbar \omega_{\perp})$, where $a_s$ is the s-wave
scattering length, $M$ is the particle mass, and $\omega_{\perp}$ is
the trap frequency. The LLL approximation requires $\lambda \ll 1$,
while the parameters of Ref.  \cite{Ketterle} correspond to at least
$\lambda \in 1-100$. Under such conditions, we expect Eq.
(\ref{traphamilton}) to be adequate.  (See also caption of Fig.
\ref{fig:vortexpositions}).

In the problem of vortex matter in a trap we encounter two specific
circumstances, namely a finite-size situation and an inhomogeneous
density profile. We begin by examining the consequences of finite
size, by studying the system in a cylindrical container {\it with a
  uniform $P_{ij}$}. Such a situation is indeed relevant for the
physics of liquid ${}^4$He. Vortex orderings in such a geometry at
zero temperature were studied in \cite{zipf}, however the VLL melting
for this case was addressed only for a planar geometry (see e.g.
\cite{2dm}). Since the melting process in three dimensional vortex
matter is very different from that in two dimensions due to the
importance of vortex line bending, this problem warrants careful
consideration.

\begin{figure}[htbp]
  \centerline{\hbox{\includegraphics[width=85mm]{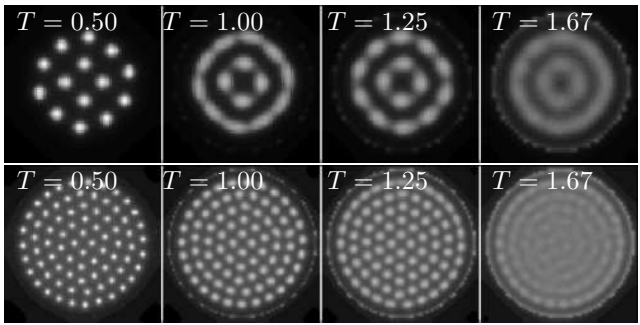}}}
  \caption{$xy$ positions of vortices in a cylindrical container
    integrated over $z$-direction, and averaged over every tenth of a
    total of $5 \cdot 10^5$ MC sweeps. Top and bottom rows have $L =
    36$ and $L=72$, respectively. At $T=0.5, 1.0$, we discern circular
    ordering close to the cylinder wall combined with a hexagonally
    ordered state closer to the center. At $T=1.25, 1.67$ we observe
    dominance of angular fluctuations closest to the edge. }
    \label{fig:hardboundary}
\end{figure}

Fig. \ref{fig:hardboundary} shows the results of simulations of vortex
matter in a cylindrical container with $P_{ij}=1, r_{ij} \leq R$. At
low temperatures, the simulations reproduce orderings with circular
distortions of VLL near the container wall, as predicted for ${}^4$He
in a zero-temperature treatment of the problem \cite{zipf}. For a
large number of vortices the system reacquires the hexagonal lattice
symmetry away from the wall, see Fig. \ref{fig:hardboundary} (bottom
row). Increasing the temperature in the case of small number of
vortices (top row of Fig. \ref{fig:hardboundary}) the dominant vortex
fluctuations are associated with angular displacements, while radially
the vortex density remains ordered. For a larger number of vortices
(bottom row of Fig. \ref{fig:hardboundary}) we find dominance of
angular fluctuations only for the vortices situated close to container
wall, while the center of the system does not display this phenomenon.
The crossover to a uniformly molten vortex system occurs in both cases
only at a higher temperature. The two-step thermal crossover in the
vortex pattern we find is analogous to that in two dimensions where
vortices are point-like objects (see e.g.  \cite{2dm}).  There is,
however, a principal difference in our case, since in three dimensions
the VLL melting is accompanied by significant vortex bending
fluctuations.
  
Let us now turn to the case of a harmonic trap (Eq.
\ref{traphamilton}). In a uniform and infinite system, the
fluctuations can cause either VLL melting via a first-order phase
transition or a second order transition associated with a thermally
induced proliferation of closed vortex loops near the critical
temperature where the vortices loose their line-tension
\cite{tesanovic,sudbo}.  In a finite-size inhomogeneous system, the
situation is different. A density gradient in a trap may effectively
be viewed as a temperature gradient in a uniform system. It is clear
that for low, but finite, temperatures there will be a finite area
near the edge of the cloud which effectively would be at a high enough
temperature to feature an annulus of tension-less tangle of vortices.
This is a phase where the vortex-line tension (free energy per unit
length) has vanished through the proliferation of vortex loops
\cite{sudbo}. {\it The boundary where this tangle sets in, marks the
  true boundary of the BEC}. An issue to be adressed is whether we
encounter an appreciable VL (i.e.  tensionful but disordered vortex
state) layer between the tension-less vortex tangle and the VLL.
 
\begin{figure}[htbp]
  \centerline{\hbox{\includegraphics[width=85mm]{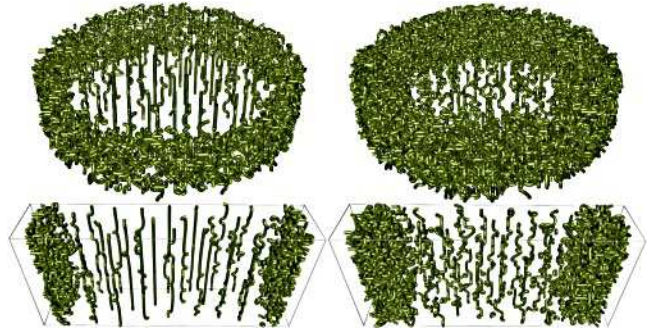}}}
  \caption{(Color online) Snapshots of vortex configurations in a
    rotating trapped BEC at $T = 0.5$ (left figures) and at $T=1.0$
    (right figures). The top row shows a selection of $16$ out of $72$
    layers in $z$ direction.  The bottom row shows smaller selections
    in the $xy$ plane, but $32$ out of $72$ layers in $z$ direction.
    Fluctuations are minimal in the trap center, and increase towards
    the edge of the trap. A distinct front separating regions of
    ordered and disordered vortices is easily identified.}
    \label{fig:3dsnapshot}
\end{figure}

\begin{figure}[htbp]
    \centerline{\hbox{\includegraphics[width=85mm]{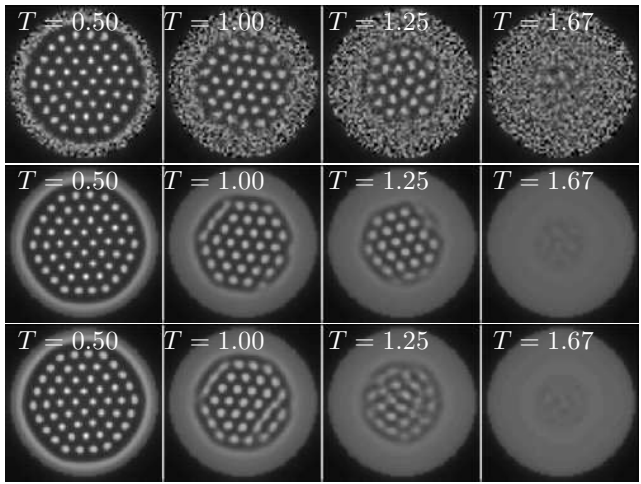}}}
    \caption{$xy$-positions of vortices in a trapped BEC integrated
      over $z$-direction.  Top row are snapshots, while middle and
      bottom rows are averages of $10^5$ and $5 \cdot 10^5$ MC sweeps,
      respectively.  Every tenth configuration has been sampled. This
      provides information on the stability of the ordered region and
      the evolution of the disordered region as $T$ varies.
      Intervortex distance $2 r_0 = 2 \sqrt{\hbar/M \Omega}$ and
      healing length $\xi = \hbar/\sqrt{2 M g_{2D} n} \leq c a/2$,
      where $a$ is lattice constant and $c$ is a constant of order,
      but less than, unity. Hence, $2 \hbar \Omega/g_{2D} n \leq
      c^2/9$, which according to Coddington {\it et al.} \cite{exp}
      validates our approach. }
\label{fig:vortexpositions}
\end{figure}

Fig. \ref{fig:3dsnapshot} shows snapshots of vortex configurations in
the model Eq. \ref{traphamilton} generated by MC simulations at
$T=0.5$ and $T=1.0$.  For better visualization we choose the vortex
radius to be $0.4$ of the grid spacing so it should \emph{not} be
associated with the core size. The sharp bends at short length scales
result from the presence of a numerical grid.  Nonetheless, this model
has proved to be accurate for describing vortex fluctuations at scales
larger than the grid spacing \cite{sudbo,hetzel,tachiki}.

We next locate the vortex liquid layer between the ordered vortex
state and the tension-less vortex tangle closest to the edge of the
trap. To this end, we take snapshots of vortex matter at different
temperatures, in each case integrating over the $z$-direction.  In a
resulting image, straight vortex lines will be seen as bright spots
while bent vortices will be seen as smeared out spots. This may be
related to experiments, where at least for non-equilibrated vortex
systems the $z$-integration renders vortices essentially
indistinguishable \cite{equilib,bretin}.  The results are given in the
upper row of Fig. \ref{fig:vortexpositions} for different
temperatures.  There we can identify regions of rather straight and
ordered vortex lines and a smeared region.
To obtain further insight into the vortex matter in this case, we also
perform a \emph{thermal averaging} as in Fig. \ref{fig:hardboundary}.
This is shown in the second and third rows in Fig.
\ref{fig:vortexpositions}. By averaging over different number of
snapshots we identify a well-defined boundary between the ordered and
disordered regions. Indeed, in a finite system, averaging will
eventually produce a complete smearing even in the center of the trap.
We observe signatures of this effect in the clear differences between
the third picture in second and third row of Fig.
\ref{fig:vortexpositions}, where averaging was made over $10^5$ and $5
\cdot 10^5$ MC sweeps, respectively. Thus, the time scale of the
fluctuations in the ordered regions are dramatically larger than those
related to the fluctuations in the disordered regions.
 
We next investigate the character of the vortex state in the
disordered region.  It is known that in the VL the helicity modulus,
or equivalently the superfluid density, is zero in any direction
\cite{tachiki,sudbo}. Monitoring of the helicity modulus could be
employed to identify a region of a possible tensionful VL in the above
pictures, as explained below.  In a trapped system the global helicity
modulus $\Upsilon_z$ \cite{sudbo} has no rigorous meaning, due to the
non-uniform $P_{ij}$. However, we introduce a modified helicity
modulus in $z$ direction, defined in a selected region between two
cylinders of radii $R_1$ and $R_2$.  We do so by applying a twist
\begin{equation}
  \label{eq:twist}
  \ve{\Delta}(r_{ij}) \equiv \ve{\Delta}_{ij} = 
  \begin{cases}
    \Delta \ve{\hat{z}} & \text{if $R_1 \le r_{ij} < R_2$},\\
    0 & \text{otherwise},
  \end{cases}
\end{equation}
to the model Eq. \ref{traphamilton} and defining the modified helicity
modulus as follows,
\begin{align}
  \label{eq:cylheli}
  \tilde{\Upsilon}_z(R_1,R_2&) \equiv 
  %% \left.\frac{\partial^2
  %%     F}{\partial \Delta^2}\right|_{\Delta = 0} = 
  \frac{1}{N^{\prime}} \left\langle {\sum}^{\prime} P_{ij}
    \cos(\theta_j - \theta_i -
    A_{ij})\right\rangle \nonumber \\
  &- \frac{1}{T N^{\prime}} \left\langle {\sum}^{\prime} \left[P_{ij}
      \sin(\theta_j - \theta_i - A_{ij})\right]^2\right\rangle.
\end{align}
Here, ${\sum}^{\prime}$ is over all links where $\ve{\Delta}_{ij}$ is
nonzero (depending on $R_1$ and $R_2$) and $N^{\prime}$ is the number
of these links.

In a uniform extended system the proliferation of vortex loops happens
via a second order phase transition \cite{sudbo}. When loops
proliferate, the condensate is destroyed. The temperature of
vortex-loop proliferation decreases with increasing rotation (see Fig.
12 in \cite{sudbo}). Alternatively, the destruction of the
phase-coherence along the $z$-axis is caused by destruction of the
lattice order via a first order transition. This scenario, if it is
realized in a trapped system, should be manifest in the {\it shape} of
the helicity modulus as the transition is approached, in that it
should be significantly different from the case without rotation.
Namely, one should see a {\it remnant of a first order phase
  transition} with a near-discontinuity in the helicity modulus,
rather than the continuous variation characterizing a second order
transition driven by a proliferation of vortex loops.  If one were to
observe no appreciable difference in the temperature dependence of the
helicity with and without rotation, one would conclude that the
demarcation line seen in the images separates an ordered region from
tension-less vortex tangle, with no discernible VL region.  This
scenario would imply a well defined and regular structure of the
boundary of the VLL.

\begin{figure}[htbp]
  \centerline{\hbox{\includegraphics[width=85mm]{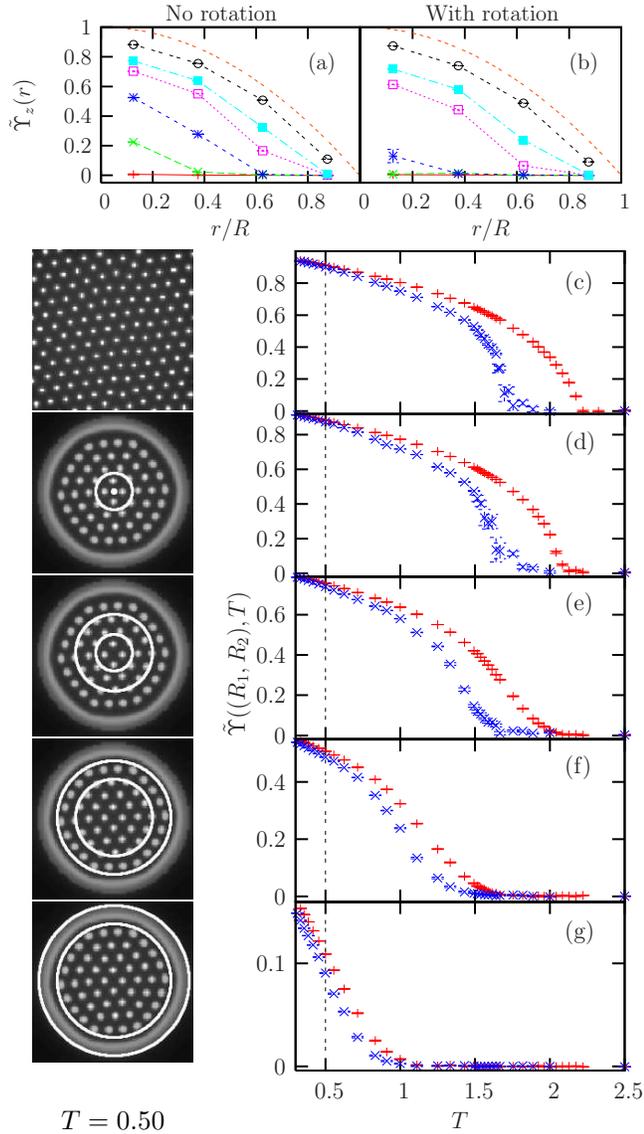}}}
  \caption{Results for $\tilde{\Upsilon}_z(R_1,R_2)$.  The two top
    panels show thermal depletion of the superfluid density in the
    model Eq.  \ref{traphamilton} ($r$ is the distance from the center
    of the trap) at the temperatures $T=2.50$ (the lowermost curve),
    $T=2.00$, $T=1.67$, $T=1.25$, $T=1.00$, $T=0.50$.  In panels
    (c)--(g), the upper curve ($+$) is the helicity modulus without
    rotation, while the lower curve ($\times$) the helicity modulus
    with rotation-induced vortices with filling fraction $f = 1/36$ as
    functions of temperature. Panel (c) shows $\Upsilon_z$ for a cubic
    uniform system with periodic boundary conditions.  The upper curve
    ($+$) has the properties of a second order transition (helicity
    modulus vanishes because of vortex-loop proliferation), whereas
    the lower curve ($\times$) has the finite-size appearance of a
    first order transition (suggesting that the helicity modulus
    vanishes because of vortex lattice melting) \cite{tachiki,sudbo}.
    The remaining panels (d)-(g) show $\tilde{\Upsilon}_z(0,R/4)$,
    $\tilde{\Upsilon}_z(R/4,2R/4)$, $\tilde{\Upsilon}_z(2R/4,3R/4)$,
    and $\tilde{\Upsilon}_z(3R/4,R)$, respectively. The vortex plots
    on the left (obtained at $T=0.50$) defines the radii $R_1$ and
    $R_2$ as white circles. Taking parameters from Ref.
    \onlinecite{Ketterle}, and using $\Omega = (h/M) N_v/2 \pi R^2$
    \cite{Ketterle} where $N_v$ is the number of vortices in the trap,
    we find $\Omega \sim 100 \rm{Hz}$. Since $\omega_{\perp} \sim 500
    \rm{Hz}$ \cite{Ketterle}, this puts us well outside the LLL
    regime.}
      \label{fig:helimod}
\end{figure}

The results for $\tilde{\Upsilon}_z(R_1,R_2)$ are shown in Fig.
\ref{fig:helimod}.  These measurements indeed show that the presence
of a rotation significantly reduces the temperature at which
$\tilde{\Upsilon}_z$ vanishes. This reduction relative to the case of
no rotation decreases with increasing $R_1,R_2$.  That is, panel (d)
is similar to panel (c) (no trap), whereas in panel (g) there is
little difference between $\tilde{\Upsilon}_z(R_1,R_2)$ with and
without rotation.  Thus, for the latter case the presence of vortices
essentially does not influence $\tilde{\Upsilon}_z$, and the
destruction of superfluid density is driven by the proliferation of
vortex loops.  Panel (g) is connected to the three leftmost panels in
Fig. \ref{fig:vortexpositions} in the following sense. The distinct
demarcation line in these leftmost images separates a vortex solid
from a tension-less vortex tangle with no visible tensionful VL
region.  This is consistent with the experiments showing a very
regular edge structure for systems with large number of vortices
\cite{exp,Ketterle}. Note also the absence of circular distortion for
the vortices at the edge of the system.  On the other hand, by
comparing panels (c) and (d), we see that $\Upsilon_z$ and
$\tilde{\Upsilon}_z$ are quite similar.  This points toward a
possibility of the existence of a tensionful VL phase close to the
center of the trap.

We have considered vortex matter in the model Eq.  \ref{traphamilton}
with uniform and nonuniform density. The uniform case features
dominant angular vortex fluctuations near the wall of the cylinder.
Vortex matter in a trapped BEC is more complicated, due to density
gradients.  We have identified a number of inhomogenous vortex states.
Notabley, we find a direct crossover from a vortex solid to
tension-less vortex tangle with no discernible intermediate tensionful
vortex liquid at low temperatures as the edge of the trap is
approached.  This explains very regular edges of VLL at finite
temperatures in many experiments. At higher temperatures, a possible
tensionful vortex liquid state located between a vortex solid at the
center and a tension-less vortex tangle closer to the edge is
identified. Our simulations indicate strong bending fluctuations of
vortices in this region which may obscure its visibility in
experiments. An experimental observation of the discrepancy between
visible vortex numbers and rotation frequency (as we predict for
disordered vortex states) has in fact been observed, albeit in an
anharmonic trap \cite{bretin}.

This work was supported by the Research Council of Norway, Grant Nos.
158518/431, 158547/431 (NANOMAT), 167498/V30 (STORFORSK), the National
Science Foundation, Grant No. DMR-0302347, Nordforsk Network on
Low-Dimensional Physics. Discussions and correspondence with J.
Dalibard, A. K. Nguyen, V. Schweikhard, E.  Sm\o rgrav, M. Zwierlein
and especially with Erich J. Mueller, are gratefully acknowledged.


\begin{thebibliography}{99}
\bibitem{details}
  At least $10^5$ MC sweeps were discarded for equilibration at each
  temperature and up to $5 \cdot 10^5$ sweeps were used for
  calculating averages.
\end{thebibliography}
\end{document}